\documentclass[letterpaper, 10 pt, conference]{ieeeconf}  

\IEEEoverridecommandlockouts                              
\usepackage{xcolor}
\usepackage{subfigure}
\usepackage{graphicx}
\usepackage[frozencache=true,cachedir=minted-cache]{minted} 
\usepackage{listings}
\setminted{fontsize=\footnotesize,baselinestretch=.97,linenos,frame=lines,xleftmargin=6pt,numbersep=3pt,mathescape=true,escapeinside=^^,bgcolor=bg}
\definecolor{bg}{rgb}{0.97,0.97,0.97}
\newminted{chapel}{}

\title{\LARGE \bf
Scaling Shared-Memory Data Structures as Distributed Global-View Data Structures in the Partitioned Global Address Space model}

\author{Garvit Dewan, Indian Institute of Technology Roorkee, gdewan@cs.iitr.ac.in \\ Louis Jenkins, University of Rochester, louis.jenkins@rochester.edu\thanks{Special thanks to Cray, a Hewlett-Packard Enterprise Company, for providing access to the compute resources utilized during development.}}

\begin{document}

\maketitle
\thispagestyle{empty}
\pagestyle{empty}

\begin{abstract}

The Partitioned Global Address Space (PGAS), a memory model in which the global address space is explicitly partitioned across compute nodes in a cluster, strives to bridge the gap between shared-memory and distributed-memory programming. To further bridge this gap, there has been an adoption of global-view distributed data structures, such as `global arrays' or `distributed arrays’. This work presents the Distributed Interlocked Hash Table (DIHT), a global-view distributed map data structure, which is inspired by the shared-memory Interlocked Hash Table (IHT).
At 64-nodes with 44-cores per node, the DIHT provides up to 110x the performance of the state-of-the-art Chapel's HashedDist. This work also demonstrates how shared-memory data structures can be modified to scale in distributed memory.  

\end{abstract}

\section{INTRODUCTION}

Distributed data structures are significantly challenging to design and validate as correct~\cite{kshemkalyani2011distributed}. The distributed-memory architecture is the primary source of this difficulty: inter-node communication is expensive and can degrade performance. This is exacerbated by concurrency on each node, which has its own set of challenges~\cite{mAlgorithmsScalableSynchronization1991}. An increased number of threads in the system may cause more memory contention, bringing down performance~\cite{robison2007too}. Algorithmic improvements aimed at improving performance frequently make designing and verifying an appropriate data structure implementation more difficult~\cite{gramoli}.

Distributed data structures have typically been designed and implemented with a ``local-view" of parallel computation in mind~\cite{barrett2008exploring}, requiring the code developer to explicitly manage both the interaction of parallel processes and the overall data layout. In contrast, PGAS languages such as Chapel provide a ``global-view" of the computation and associated data to make it easier to write code for parallel processing architectures. 
The overarching goal is to combine a global view of the program with the tools required for injecting high-level programmer ``intent" that the compiler cannot easily detect in more traditional programming models. 
The global-view model, in particular, eliminates the syntactic distinction between local and remote memory access found in local-view languages~\cite{hayashiLLVMbasedCommunicationOptimizations2015}. 

The recent effort of the AtomicObjects library in Chapel~\cite{epochManager} enabled distributed atomic operations on arbitrary objects.
Memory reclamation is at the basis of many problems in the design of high-performance data structures; hence caution should be given when reclaiming memory that may be accessed concurrently in a language without garbage collection. Efforts such as EpochManager~\cite{epochManager} have bridged that gap by providing a scalable, concurrent-safe distributed non-blocking memory reclamation system, thus enabling the implementation of more complex data structures in the distributed memory.
All of these efforts have paved the way for the incorporation of several shared-memory concepts into distributed memory.

This paper demonstrates how a shared-memory data structure can be made to scale as a global-view data structure in distributed memory by using privatization to distribute the data structure and EpochManager \cite{epochManager} for memory reclamation. We present the Distributed Interlocked Hash Table (DIHT), a data structure inspired by the Interlocked Hash Table~\cite{jenkins2017redesigning} and designed for distributed memory using the concepts discussed above. DIHT is a distributed hash table that allows for simultaneous insertions, deletions, and lookups. It, like the IHT, allows programmers to write large critical sections over a single map element, or even multiple elements. DIHT also supports parallel iteration in addition to serial iteration.

\section{Background and Related Work}
This work has been primary enabled by AtomicObjects library~\cite{epochManager}, which enables atomic operations on arbitrary class objects, both in shared and distributed memory; and EpochManager~\cite{epochManager}, a concurrency-safe memory reclamation system designed for distributed memory. In this section we briefly discuss the functioning of EpochManager and the Interlocked Hash Table.

\subsection{Epoch Manager}
EpochManager is a distributed non-blocking concurrent-safe memory reclamation system that is based on epoch-based reclamation (EBR)~\cite{fraser2004practical}. EpochManager is based on the concept of limbo lists, which hold objects marked for deletion in the limbo-list of the corresponding epoch and delete them when it is safe to do so. EpochManager is designed in such a way that it can be paired with distributed implementations of data structures for efficient memory management, allowing for distributed global-view programming.  All objects in the limbo list are deleted at once, rather than incrementally.

The EpochManager generates a collection of tokens, which are class instances that keep track of the epoch in which a task is currently engaged. Before a task may access a data structure protected by the EpochManager's epoch-based reclamation, it must first register and obtain one of these tokens. As an optimization, a token can be used to perform multiple operations in the same task. Once registered, the token is not yet in an epoch. Tokens must be pinned and unpinned in the same way that they must be registered and unregistered. When an object is to be deleted, it is always added to the token's current epoch. Global-epoch is advanced when no thread is active in any previous epochs, allowing for the reclamation of objects marked for deletion in previous epochs.

\subsection{Interlocked Hash Table}
The Interlocked Hash Table (IHT) is a lock-based map that supports concurrent insertions, deletions, and lookups. Initially designed for the Go programming language, the IHT uses a speculative traverse of a fixed-max-depth tree of intermediary nodes to obtain exactly one lock every insert/remove/update/lookup operation. To avoid deadlock, the IHT implements a novel optimistic locking strategy and supports large critical sections that access a single IHT element. Even when a high number of threads are executing simultaneous random accesses to a small map, the net result is low contention since locks and data are co-located. IHT's high scalability motivated the authors to experiment its adaptation in distributed memory.

\section{Design \& Implementation}

The DIHT, like the RCUArray~\cite{jenkinsRCUArrayRCULikeParallelSafe2018}, addresses three major challenges: (1) parallel-safe memory reclamation; (2) operation concurrency; and (3) distribution across numerous cluster nodes. Analogous to the IHT, it makes use of two major data types: \texttt{ElementLists}, which hold key/value pairs, and \texttt{PointerLists}, which hold references to either \texttt{ElementLists} or other \texttt{PointerLists}. Table~\ref{tab:base_objects} shows fields from both of these data types. The \texttt{PointerList} and \texttt{ElementList} classes both derive from the Base class. The Base class's Lock field is an atomic int that, in addition to ensuring mutual exclusion in \texttt{ElementLists}, also offers knowledge to discriminate between \texttt{ElementList} and \texttt{PointerList} at runtime. \texttt{PointerLists} are never locked. The possible lock states, which are analogous to IHT, are shown below:
\begin{itemize}
    \item $e_avail$ - An unlocked \texttt{ElementList}.
    \item $e_lock$ - A locked \texttt{ElementList}.
    \item $p_list$ - A \texttt{PointerList}.
    \item $GARBAGE$ - \texttt{ElementList} marked for deletion. 
\end{itemize}

The distributed map is defined by its root, which is a \texttt{PointerList} with its buckets block distributed across the cluster's nodes, and three constants: \textit{ROOT\_BUCKETS\_SIZE}, the size of the buckets array in the root \texttt{PointerList}; \textit{BUCKET\_NUM\_ELEMENTS}, the maximum number of elements in any \texttt{ElementList}; and \textit{BUFFER\_SIZE}, the size of the aggregation buffer, which is discussed later. 

The distributed map, like the IHT, can grow but not shrink. \texttt{PointerLists} cannot be deleted, but \texttt{ElementLists} can.  As a result, once a bucket refers to a \texttt{PointerList}, it is immutable.

\begin{table}
    \centering
    \caption{\texttt{ElementList} and \texttt{PointerList} Fields}
    \label{tab:base_objects}
    \begin{tabular}{l l l}
        \multicolumn{3}{l}{\textbf{Fields of Base}} \\
        lock & : atomic int & // Holds the lock value \\
        parent & : Base & // parent bucket \\
        
        \multicolumn{3}{l}{\textbf{Fields of Bucket (ElementList) (extends Base):}} \\
        count & : Integer & // number of active elements \\
        keys & : keyType[] & // keys stored in ElementList \\
        values & : valType[] & // values stored in ElementList \\
        
        \multicolumn{3}{l}{\textbf{Fields of Buckets (Pointerlist) (extends Base):}} \\
        bucketsDom & : domain & // Domain of linked buckets \\
        buckets & : AtomicObject[] & // Pointers of linked buckets \\
        seed & : Integer & // Seed for hashing \\
    \end{tabular}
\end{table}

The data type used to define the map is shown in table~\ref{tab:map_struct}. The data type is \textit{privatized}, which means that an instance is allocated on each node to which all accesses are directed, resulting in zero communication when acquiring the privatized instance. This is accomplished through \textit{remote-value forwarding} and \textit{record-wrapping}, in which a record containing data required to look up the instance itself is based on value rather than reference. This results in a massive speedup since replication across locales eliminates all unnecessary communication and allows for data caching or even the retention of locale-specific instances of data. The record-wrapping eliminates the need for an additional round-trip communication to obtain the metadata required to locate the privatized instance. In practice, the authors have observed that this allows distributed objects to no longer be communication-bound, allowing for truly scalable algorithms. The \textit{PID} field, also known as the privatization id, is a descriptor that is used to access the privatized instance assigned to each node. \textit{KeyType} is the key data type; \textit{ValType} is the value data type. \textit{RootArray} forms the root of the map, block distributed across all the nodes. \textit{Aggregator} is an aggregation buffer that is used for aggregating and executing asynchronous operations. \textit{Manager} is an EpochManager instance that is used for memory reclamation. Two different random number generators are used: \textit{IterRNG} for map iterations and \textit{SeedRNG} for bucket seed generation.

\begin{table}
    \centering
    \caption{DIHT Fields}
    \label{tab:map_struct}
    \begin{tabular}{l l l}
        \multicolumn{3}{l}{\textbf{Constants}} \\
        rootBucketSize & : Integer & // root PointerList size \\
        bucketNumElements & : Integer & // size of an ElementList \\
        bufferSize & : Integer & // aggregator buffer size \\
        rootSeed & : Integer & // root seed for hashing \\
        
        \multicolumn{3}{l}{\textbf{Fields of DistributedMapImpl}} \\
        pid & : Integer & // privatization id \\
        rootBuckets & : RootBuckets & // distributed root \\
        aggregator & : Aggregator & // aggregator for async ops \\
        manager & : EpochManager & // for memory reclamation \\
        seedRNG & : RandomStream & // RNG for PList seeds \\
        iterRNG & : RandomStream & // RNG for iteration \\
        
        \multicolumn{3}{l}{\textbf{Fields of RootBuckets}} \\
        dom & : domain & // block-distributed domain \\
        buckets & : AtomicObject[] & // root buckets \\
    \end{tabular}
\end{table}



\subsection{Behavior}
With the distributed map design in place, we can now talk about how the distributed map behaves. There are numerous techniques to distribute components over map nodes. One solution may be to locally store each element on the node on which the insert operation is called, resulting in faster inserts since it is always a local operation but slower look-ups and removals. Because some nodes may do more inserts than others, local insertions may cause the map to be spread unevenly. This, in turn, would increase traffic on these nodes for look-ups. Hashing is a solution to this problem: for each element, a rootHash of its key is calculated, and the node is chosen based on this hash value. Insertions, look-ups, and deletions are then defined as a Remote Procedure Call (RPC) on the appropriate node. In this strategy, all operations may include a network call, and there may not be a single fast operation. However, given the typical load of 80\% lookups and 20\% insertions and deletions, this is a preferable method because evenly spread data ensures no extra traffic on any nodes. In the implementation of this distributed map, the latter strategy was adopted.

\subsection{Algorithm Correctness}
With the behavior of the distributed map sketched, we now present the algorithms. The insert, lookup, and erase operations follow a familiar pattern: find the node on which the element is located, make an RPC call to the respective node to traverse the tree, lock the respective \texttt{ElementList}, perform the operation and unlock the \texttt{ElementList}. Synchronized insert, lookup, and erase are presented in listings~\ref{lst:insert},~\ref{lst:lookup} and~\ref{lst:remove} respectively. The algorithmic correctness of all these functions can be established based on IHT's algorithms' correctness. Each operation obtains no more than one lock, eliminating the possibility of a deadlock.

The traversal and expansion algorithm has been encapsulated in the getEList function, which is is similar to IHT's getEList algorithm combined with epoch manager to reclaim \texttt{ElementLists} rehashed into \texttt{PointerLists}. The function is always executed locally on each node. Beginning with the root \texttt{PointerLists}, first the index is calculated based on the rootHash. If the bucket at this index is \texttt{nil} and the operation is not an insert operation, then \texttt{nil} is returned. Else, a new \texttt{ElementList} is created and compare-and-swapped to the \texttt{PointerList}'s bucket at the calculated index. If successful, the \texttt{ElementList} is returned, else the \texttt{ElementList} is marked for deletion by calling \textit{token.tryReclaim(elist)}. If the bucket is not nil but a \texttt{PointerList}, we recurse over it. If the bucket is an \texttt{ElementList}, we try to acquire its lock. If successful, the \texttt{ElementList} is returned if it is not full, else it is rehashed into a \texttt{PointerList} with the old \texttt{ElementList}'s elements inserted as new \texttt{ElementLists} into the \texttt{PointerList}. The old \texttt{ElementList} is marked for deletion. We then recurse over this \texttt{PointerList}. If \texttt{ElementList} lock could not be acquired, the thread backs off and tries the entire procedure.

\begin{listing}
 \begin{minted}[breaklines=true,fontsize=\scriptsize,linenos=false]{chapel}
proc insert(key, val, tok) {
  tok.pin(); // pin the epoch manager
  // Get index of target locale
  const idx = this.getIdx(key);
  var _pid = pid;
  // the RPC call
  on rootBuckets[idx].locale {
    // get privatized copy of instance
    var _this = chpl_getPrivatizedCopy(this.type, _pid);
    _this.insertLocal(key, val, tok);
  }
  tok.unpin(); // unpin the epoch manager
}

// Helper function for node-local insert
proc insertLocal(key, val, tok) {
  var elist = getEList(key, tok);
  for i in 1..BUCKET_NUM_ELEMS {
    // if key already exists, update value
    if (elist.keys[i] == key) {
      elist.values[i] = val;
      elist.lock.write(E_AVAIL);
      return;
    }
  }
  // 1-indexed
  elist.count += 1;
  elist.keys[count] = key;
  elist.values[count] = val;
  elist.lock.write(E_AVAIL); // release the lock
}
\end{minted}
\caption{Insert operation}
\label{lst:insert}
\end{listing}

\begin{listing}
 \begin{minted}[breaklines=true,fontsize=\scriptsize,linenos=false]{chapel}
proc find(key, tok) : (bool, valType) {
  var found = false; // true if key was found
  var retVal : valType?; // stores value
  tok.pin(); // pin the epoch manager
  // Get index of target locale
  const idx = this.getIdx(key);
  var _pid = pid;
  // the RPC call
  on rootBuckets[idx].locale {
    // get privatized copy of instance
    var _this = chpl_getPrivatizedCopy(this.type, _pid);
    (found, retVal) = _this.findLocal(key, val, tok);
  }
  tok.unpin(); // unpin the epoch manager
  return (found, retVal);
}

// Helper function for node-local find
proc findLocal(key, tok) {
  var elist = getEList(key, tok);
  var retVal : valType? = nil;
  var found = false;
  if (elist != nil) {
    for i in 1..BUCKET_NUM_ELEMS {
      if (elist.keys[i] == key) {
        found = true;
        retVal = elist.values[i];
        break;
      }
    }
    // release the lock
    elist.lock.write(E_AVAIL);
  }
  return (found, retVal);
}
\end{minted}
\caption{Find operation}
\label{lst:lookup}
\end{listing}

\begin{listing}
 \begin{minted}[breaklines=true,fontsize=\scriptsize,linenos=false]{chapel}
proc erase(key, tok) {
  tok.pin(); // pin the epoch manager
  // Get index of target locale
  const idx = this.getIdx(key);
  var _pid = pid;
  // the RPC call
  on rootBuckets[idx].locale {
    // get privatized copy of instance
    var _this = chpl_getPrivatizedCopy(this.type, _pid);
    _this.eraseLocal(key, val, tok);
  }
  tok.unpin(); // unpin the epoch manager
}

// Helper function for node-local erase
proc eraseLocal(key, tok) {
  var elist = getEList(key, tok);
  if (elist == nil) then return;
  for i in 1..BUCKET_NUM_ELEMS {
    if (elist.keys[i] == key) {
    // lazy delete by swapping with last element
      elist.keys[i] = elist.keys[count];
      elist.values[i] = elist.values[count];
      elist.count -= 1;
      break;
    }
  }
  // if elist is empty, delete it
  if (elist.count == 0) {
    // Marking the elist for deletion
    elist.lock.write(GARBAGE);
    tok.deferDelete(elist);
  } else {
    // release the lock
    elist.lock.write(E_AVAIL);
  }
}
\end{minted}
\caption{Remove operation}
\label{lst:remove}
\end{listing}



\subsection{Aggregation}
Because elements are evenly distributed across all nodes in the cluster, the likelihood that an operation will be performed remotely increases as the number of nodes increases. As a result, while increasing the number of nodes in a cluster increases the number of threads available and thus the number of operations that can be executed in parallel, it also increases the amount of inter-node communication, which is expensive and reduces performance~\cite{kshemkalyani2011distributed}. The problem is solved by executing the operations asynchronously by aggregating the operations based on the node on which the operation must be executed. Multiple RPCs between the same sender and receiver nodes are batched into fewer calls, lowering the total number of remote calls. Local operations are never aggregated and are always carried out immediately. In distributed-map, aggregation is achieved using the Chapel Aggregation Library~\cite{jenkinsChapelAggregationLibrary2018}, which provides an aggregator with a fixed size buffer. \textit{insertAsync}, \textit{findAsync}, and \textit{eraseAsync} are asynchronous variations of the Insert, Lookup, and Erase functions that are shown in listing~\ref{lst:asyncOp}. The EmptyBuffer algorithm is also presented in listing~\ref{lst:emptyBuffer}.

Each function first determines the node on which the operation will be performed. If the operation is to be performed locally, it is carried out immediately. Otherwise, the operation is added to a local instance of aggregator, along with the locale id on which the operation is to be performed. While \textit{insertAsync} and \textit{eraseAsync} do not return anything for optimization purposes, \textit{findAsync} returns a future object, which is updated to reflect the status of the find operation once it is executed along with the return value if the operation succeeded. The operations are asynchronously pushed to the corresponding locale for execution via an RPC when the aggregation buffer for a node's locale is filled. The buffer is data-parallel iterated, and the instructions are executed by invoking their corresponding local routines. If \textit{findAsync} calls are detected, a remote buffer is created to store the successful operations and their return values. When all operations have been executed, this buffer is returned to the caller node, and the futures of the respective \textit{findAsync} calls are updated from it. This approach assists in limiting \textit{BUFFER\_SIZE} remote calls to a maximum of two remote calls, hence lowering inter-node communication dramatically. This procedure is encapsulated in EmptyBuffer function displayed in listing~\ref{lst:emptyBuffer}.

\begin{listing}
 \begin{minted}[breaklines=true,fontsize=\scriptsize,linenos=false]{chapel}
proc findAsync(key : keyType, tok) {
  const idx = this.getIdx(key);
  var future = new MapFuture(valType);
  // if element exists locally, execute immediately
  if here.id == rootBuckets[idx].locale.id {
    tok.pin();
    var (found, val) = findLocal(key, tok);
    tok.unpin();
    // if key found, update value in future
    if found then future.success(val);
    else future.fail();
  } else {
    // aggregate onto the target locale's buffer
    var buff = aggregator.aggregate((MapAction.find, key, future), rootBuckets[idx].locale);
    // if buffer is full, empty it by executing all its operations on remote node
    if buff != nil {
      begin emptyBuffer(buff, rootBuckets[idx].locale);
    }
  }
  return future;
}
\end{minted}
\caption{Async Operation}
\label{lst:asyncOp}
\end{listing}

\begin{listing}
 \begin{minted}[breaklines=true,fontsize=\scriptsize,linenos=false]{chapel}
proc emptyBuffer(buffer, loc) {
  var _pid = pid;
  // get buffer contents
  const buff = buffer.getArray();
  buffer.done(); // mark buffer empty
// tuple mapped to buffer contents. Stores if the
// operation is a find operation, and its
// corresponding return values
  var findBuff : [buff.domain] (bool, bool, valType?);
  var findOpExists = false;
  on loc {
    var _this = chpl_getPrivatizedCopy(this.type, _pid); // get privatized copy
    (findOpExists, findBuff) = _this.emptyBufferHelper(buff);
  }
  // if find operations exist, update their futures
  if findOpExists {
    // data parallel loop
    forall i in findBuff.domain {
      if (findBuff[i].isFindOp) {
        if findBuff[i].success {
          retVal = findBuff[i].retVal;
          buff[i].future.success(retVal);
        } else {
          buff[i].future.fail();
        }
      }
    }
  }
}

proc emptyBufferHelper(buff) {
  var findBuff : [buff.domain] (bool, bool, valType?);
  var findOpExists = false;
  // data parallel loop with EpochManager token per thread 
  forall  i in buff.domain with (var tok = this.getToken(), ref findOpExists) {
    tok.pin();
    const action = buff[i].action;
    const key = buff[i].key;
    const val = buff[i].val;
  // switch case to call respective node-local operation
    select action {
      when MapAction.insert {
        this.insertLocal(key, val, tok);
      }
      when MapAction.find {
        findOpExists = true;
        (success, retVal) = this.findLocal(key, tok);
        findBuff[i] = (true, success, retVal);
      }
      when MapAction.erase {
        this.eraseLocal(key, tok);
      }
    }
    tok.unpin();
  }
  return (findOpExists, findBuff);
}
\end{minted}
\caption{EmptyBuffer}
\label{lst:emptyBuffer}
\end{listing}

\subsection{Iteration}

We present two iteration algorithms: serial iteration and parallel iteration. To iterate across the map serially, we start by picking a bucket at random from the local node's root \texttt{PointerList} and iterating over its collection of buckets. We roughly follow getElist's behavior for each bucket: if a bucket is \texttt{nil}, it is skipped. If a bucket is an \texttt{ElementList}, we lock it and iterate through all its elements. The lock is released once all of the elements have been iterated. If it is a \texttt{PointerList}, we recurse into it by selecting a bucket at random and repeating this process. A lazy list of \texttt{ElementList} is maintained to reduce contention over a bucket, which is local to the node. References to locked \texttt{ElementLists} are stored inside this lazy list, which are later revisited. If the \texttt{ElementList} lock is available, we iterate through its elements and remove the reference from the lazy list. If the \texttt{ElementList} lock is still unavailable, we select the following reference from the lazy list. If the \texttt{ElementList} is rehashed into a \texttt{PointerList}, the same recursive procedure is performed on the \texttt{PointerList}'s buckets. If none of the references were found unlocked, the iterator returns to the lazy list and retries iterating the references. This procedure is repeated for each root \texttt{PointerList} on the local node. Remote procedure calls are used to repeat the same procedure on each node. It is worth noting that only one lock is held at a time in this algorithm, removing the possibility of a deadlock. Multiple iterations are unlikely to iterate over the map in the same order and hence unlikely to content on the same \texttt{ElementList} at the same time, thanks to the randomized iteration.

Listing~\ref{lst:paralleliter} displays the parallel iteration algorithm. The parallel iteration algorithm employs two lock-free queues~\cite{michaelSimpleFastPractical1996} per node: \textit{WorkList}, which keeps track of all \texttt{PointerLists} on the node that need to be recursively traversed, and \textit{DeferList}, which keeps track of \texttt{ElementLists} that are deferred either because they are locked or because they are being rehashed. The algorithm employs the \textbf{coforall} construct, which distributes the set of instructions to all nodes. We start by picking a random bucket from the root \texttt{PointerList} on each node in parallel and iterating serially over their entire sets of buckets with respect to their respective nodes. If a bucket is \texttt{nil}, it is skipped. If it is an \texttt{ElementList}, we lock it and iterate through all its elements serially with respect to the node. The lock is released once all elements have been iterated. If the \texttt{ElementList} lock cannot be acquired, a reference to the \texttt{ElementList} is enqueued to the node-local \textit{DeferList}. If the bucket is a \texttt{PointerList}, a reference to the \texttt{PointerList} is enqueued to the node-local \textit{WorkList}. This procedure is repeated for each root bucket present on a node. Once all the root buckets have been iterated, we begin processing the elements referred to in \textit{WorkList} and \textit{DeferList}. Once again, the \textbf{coforall} construct is used, but this time to spawn the set of instructions across all threads on all nodes. First, the \textit{WorkList} is traversed. Each thread dequeues from the \textit{WorkList}, which returns a reference to a \texttt{PointerList}, and randomly iterates through its buckets. As with the case of root buckets iteration, \texttt{PointerLists} are enqueued to the \textit{WorkList}, unlocked \texttt{ElementLists} are iterated over in the same way as described earlier, and locked \texttt{ElementLists} are enqueued to the \textit{DeferList}. Once the \textit{DeferList} is emptied, the threads begin processing references in the \textit{DeferList}. If a reference from the \textit{DeferList} is a \texttt{PointerList}, it implies that the \texttt{ElementList} was rehashed into a \texttt{PointerList} and enqueued onto the \textit{WorkList}. If the reference is an unlocked \texttt{ElementList}, it is iterated the same way as described earlier. A locked \texttt{ElementList} is enqueued back to the \textit{DeferList}, and the thread backs off. The thread again tries to dequeue from the \textit{WorkList}, followed by querying the \textit{DeferList}. All threads synchronously exit when both \textit{WorkList} and \textit{DeferList} become empty. This iteration algorithm can scale to parallel-iterate elements in the map over all the threads in the cluster. 

Lock-free queues enable multiple threads to scale by spawning operations on the referenced \texttt{PointerLists} and \texttt{ElementLists} in parallel. By making these queues local to each node, any inter-node communication is eliminated. Like in the serial iteration, the randomized iteration ensures that multiple iterations are unlikely to iterate over the map in the same order and thus are unlikely to content on the same \texttt{ElementList} at the same time.

\begin{listing}
\begin{minted}[breaklines=true,fontsize=\scriptsize,linenos=false]{chapel}
iter these(param tag:iterKind) where tag == iterKind.standalone {
  var tok = this.getToken();
  tok.pin();
  var _pid = pid;
  // do on each node
  coforall loc in Locales do on loc {
    var _this = chpl_getPrivatizedCopy(_pid);
    _this.iterateEachNode();
  }
  tok.unpin();
}

proc iterateEachNode() {
  const sz = this.rootBuckets.size;
  const startRootIdx = random(sz);
  var workList = new LockFreeQueue();
  var deferredList = new LockFreeQueue();
  // serially visit all roots
  for i in 0..#sz {
    var rootIdx = (startRootIdx + i) % sz;
// if rootBucket exists on local node, iterate it
    if this.rootBuckets[rootIdx].locale == here {
      processElement(rootBuckets[rootIdx], rootIdx, workList, deferredList);
    }
  }

  // Now process workList and deferredList
  coforall tid in 1..here.maxTaskPar {
    while (true) {
      var plist = workList.dequeue();
      if (plist == nil) {
        var (plist, idx) = deferredList.dequeue();
        // Nothing to process, exit
        if (plist == nil) then break;

        // handling deferred
        processElement(buckets[idx], plist, idx, workList, deferredList);
        chpl_task_yield();
        continue;
      } else {
        // Process PList
        var startIdx = random(plist.buckets.size);
        for i in 0..(plist.buckets.size-1) {
          var idx = (startIdx + i)%plist.buckets.size;
          processElement(parent, idx, workList, deferredList);
        }
      }
    }
  }
}
// base -> base; cas; _this->this;
proc processElement(elem, parent, idx, workList, deferredList) {
  var base = elem.read();
  if (base == nil) then return;
  if (base.lock.read() == E_AVAIL && base.lock.CAS(E_AVAIL, E_LOCK)) {
    var bucket = base : Bucket;
    for j in 1..bucket.count {
      yield (bucket.keys[j], bucket.values[j]);
    }
    bucket.lock.write(E_AVAIL);
  } else if (base.lock.read() == P_INNER) {
    var buckets = base : Buckets;
    workList.enqueue(buckets);
  } else {
    deferredList.enqueue((parent, idx));
  }
}
\end{minted}
\caption{Parallel Iteration}
\label{lst:paralleliter}
\end{listing}


\section{Performance Evaluation}


In this section, we explore two performance criteria. First, we evaluate the performance of DIHT at varying key sizes, comparing the performance of synchronous and asynchronous operations against Chapel's HashedDist. Next, we evaluate the performance of both the iteration algorithms and compare with HashedDist's serial and parallel iteration performance. All experiments were conducted on a 64-node Cray XC-50 with 44 core Broadwell CPUs per node, compiled using Chapel 1.25 with the `--fast' flag to enable all compiler and backend optimizations.

\subsection{Operations performance}
\begin{figure}
    \centering
    \includegraphics[width=0.5\textwidth]{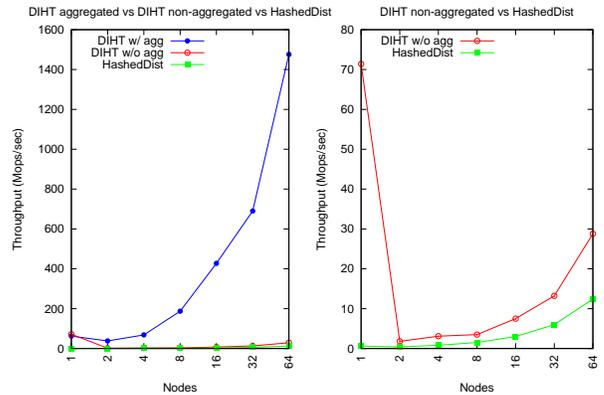}
    \caption{Microbenchmark performance}
  \label{fig:perf}
\end{figure}

For this evaluation, the DIHT was configured with 8-elements per \texttt{ElementList}. The root \texttt{PointerList} was configured to have 1024 buckets per node, while initial \texttt{PointerLists} were configured to have 1024 buckets with their size doubling at each subsequent level. Aggregation buffer size was set to 10240. Figure~\ref{fig:perf} compares the performance of DIHT's synchronous and asynchronous operations at various loads with Chapel's HashedDist. It can be observed that in the case of asynchronous operations, increasing number of find operations result in reduced throughput. This is because every buffer with a find operation makes a reverse remote call to update the future with execution status and returned value. Hence, we evaluate the map at the typical load of high reads with 80\% lookups and 20\% insertions and deletions. The benchmark can be found in listing~\ref{lst:microOp}. At read-heavy load,  DIHT without aggregation is throttled by communication, but still scales as number of nodes increase in the cluster. The throughput at 64 nodes is about half the throughput at a single node, pointing to a serious communication bottleneck. Even so, the throughput is more than Chapel's HashedDist by two times. DIHT with aggregation massively outperforms both DIHT without aggregation as well as Chapel's HashedDist by 50x and 110x respectively, and scales evenly as number of nodes increase.

\begin{listing}
\begin{minted}[breaklines=true,fontsize=\scriptsize,linenos=false]{chapel}
var map = new DistributedMap(int, int);
var tok = map.getToken();
// pre-fill the map
for i in 0..maxLimit:int do
  map.insertAsync(i, i+1, tok);
map.flushLocalBuffers(); // flush all local buffers
timer.start();
coforall loc in Locales do on loc { // on each node
  const opsperloc = N / Locales.size;
  coforall tid in 1..here.maxTaskPar { // on each thread
    var rng = new RandomStream(real);
    var keyRng = new RandomStream(int);
    const opspertask = opsperloc / here.maxTaskPar;
    var tok = map.getToken();
    for i in 1..opspertask {
    var s = rng.getNext();
      var key = keyRng.getNext(0, maxLimit:int);
      if s < 0.1 {
        map.insertAsync(key,i, tok);
      } else if s < 0.2 {
        map.eraseAsync(key, tok);
      } else {
        map.findAsync(key, tok);
      }
    }
  }
  map.flushLocalBuffers();
}
timer.stop();
var opspersec = N/timer.elapsed();
writeln(opspersec, " operations/sec");
\end{minted}
\caption{Operations Benchmark}
\label{lst:microOp}
\end{listing}

\subsection{Iteration Performance}
\begin{figure}
    \centering
    \begin{minipage}[b]{0.5\textwidth}
        \includegraphics[width=\textwidth]{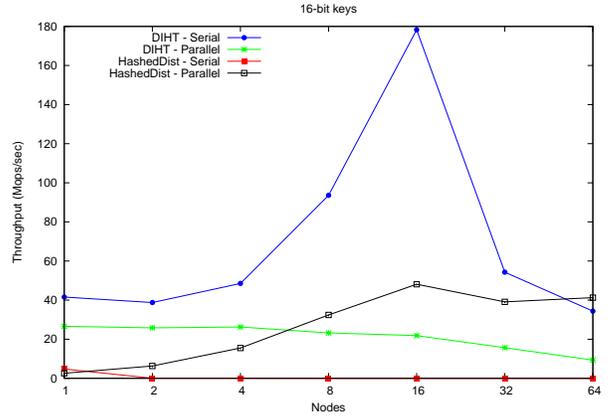}
    \end{minipage}
    \hfill
      \begin{minipage}[b]{0.5\textwidth}
        \includegraphics[width=\textwidth]{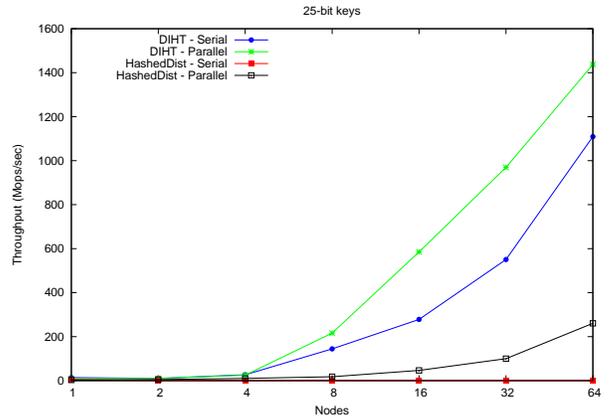}
     \end{minipage}
    \caption{Iteration microbenchmark}
    \label{fig:iteration}
\end{figure}

For evaluating iteration performance, the map is pre-filled with a fixed number of keys and then iterated. Benchmarks are performed with 65.5K keys pre-filled using 16-bit keys and 35.5M keys using 25-bit keys. Unfortunately this is the maximum number of elements that could be filled in the map without the data-structure going out-of-memory on a single node benchmark. We compare DIHT's serial-iteration and concurrent-iteration with Chapel's HashedDist. chpl\_task\_yield() is called each iteration: once per iteration in serial iteration, while a random number of times between one and ten for parallel iteration. Figure~\ref{fig:iteration} compares the performance at 16-bit keys. At 44 threads per node, the key size is too small for parallel iteration. This results in a degrading amount of throughput for DIHT's parallel iteration as number of nodes increase. On the other hand, DIHT's serial iteration results in an interesting trend, which fist scales linearly with increasing number of nodes upto 16 nodes, beyond which, the throughput begins to degrade as keys are sparsely spread across nodes. Both iteration algorithms outperform HashedDist's serial and parallel iteration algorithms initially, but at 64-nodes, HashedDist's parallel iteration performs the best.

With 25-bit keys however, we see that both serial and parallel iteration algorithms scale linearly with increasing number of nodes. DIHT's concurrent-iteration performs upto 1.4 billion operations per second at 64 nodes, outperforming HashedDist's parallel iteration by 6 times. On the other hand, DIHT's serial iteration performs comparably too, with around 1.1 billion operations per second, outperforming HashedDist's parallel iteration by over 4 times.

\begin{listing}
\begin{minted}[breaklines=true,fontsize=\scriptsize,linenos=false]{chapel}
var map = new DistributedMap(int, int);
// pre-fill the map
forall i in (-keyRange/2)..(keyRange/2 - 1) with (var tok = map.getToken()) {
  map.insertAsync(i, 0, tok);
}
map.flushLocalBuffers(); // flush all local buffers
var timer = new Timer();
timer.start();
// parallel iteration
forall i in map with (var yieldRng = new RandomStream(int)) {
  const yieldTimes = yieldRng.getNext(0, 10);
  for i in 1..yieldTimes do chpl_task_yield();
}
timer.stop();
writeln("Parallel: ", keyRange/timer.elapsed(), "ops/s.");
timer.clear();

timer.start();
// Serial iteration
for i in map {
  chpl_task_yield();
}
timer.stop();
writeln("Serial: ", keyRange/timer.elapsed(), "ops/s.");
\end{minted}
\label{lst:microIt}
\caption{Iteration Benchmark}
\end{listing}

\section{Conclusion and Future Work}

In this paper, we presented the DIHT, which is a distributed hashtable. DIHT scales in distributed memory, outperforming Chapel's HashedDist by 110x with upto 1.4 billion operations per second at 64 nodes. The work demonstrates how shared memory concepts can be adapted to distributed memory and made to scale with some optimizations. While EpochManager has been proven~\cite{epochManager} to scale in shared memory data structures such as lock-free stack~\cite{hendlerScalableLockfreeStack2004} and lock-free queue~\cite{michaelSimpleFastPractical1996}, DIHT proves the scalability of EpochManager in distributed-memory data structures.

A limitation of the current implementation is the fixed buffer size of the aggregation buffer. The buffer will wait to be filled before transferring to the targeted node. Though FlushLocalBuffers and FlushAllBuffers methods can be used to manually flush buffers, they can be expensive and non-intuitive for the programmer. A variable sized buffer would be more efficient, adjusting its size based on the load, and will result in better performance benchmarks.

DIHT, while being scalable, lacks any fault tolerance. We experimented with checkpointing schemes based on various criteria: data checkpointing on a per-node basis where in data of all the nodes in the tree is stored in files on the disk from which the map can be regenerated; and function checkpointing where in each operation to add or delete a key to the map is logged and persisted on a disk to be used later for regenrating the map. The former approach required one file pointer per \texttt{ElementList} and writing to multiple files in parallel, taking up space in memory and wasting major amount of thread execution time in disk file writes. The latter approach on the other hand may generate a large number of logs and take exorbitantly large time to process it. We also experimented with scheduled checkpointing, but ultimately the write speed on a hard drive is so low that performance degraded manifold. Persistent memory~\cite{persistent}~\cite{makalu} may offer the ideal middle ground for such checkpointing schemes, the hardware for which unfortunately could not be obtained for this work and the experimentation has thus been left as a future work.

\bibliographystyle{ieeetr}
\bibliography{references}

\end{document}